\documentclass[11pt,twoside]{article}
\usepackage{asp2010}

\resetcounters

\begin{document}

\title{Magnetic fields in nearby galaxies}
\author{Andrew Fletcher$^1$
\affil{$^1$School of Mathematics \& Statistics,
	Newcastle University,
	Newcastle upon Tyne,
	NE1 7RU, United Kingdom.}}

\begin{abstract}
Observations of synchrotron radiation and the Faraday rotation of its polarized component allow us to investigate the magnetic properties of the diffuse interstellar medium in nearby galaxies, on scales down to roughly one hundred parsecs. All disc galaxies seem to have a mean, or regular, magnetic field component that is ordered on length scales comparable to the size of the galaxy as well as a random magnetic field of comparable or greater strength. I present an overview of what is currently known observationally about galactic magnetic fields, focusing on the common features among galaxies that have been studied rather than the distinctive or unusual properties of individual galaxies. Of particular interest are the azimuthal patterns formed by regular magnetic fields and their pitch angles as these quantities can be directly related to the predictions of the mean field dynamo theory, the most promising theoretical explanation for the apparent ubiquitous presence of regular magnetic fields in disc galaxies. 
\end{abstract}

\section{Introduction}
Magnetic fields bind together the interstellar medium coupling the gas, in its many phases, to the nonthermal cosmic rays. Since all three components have roughly equal energy densities \citep{Boulares:1990} the properties and behaviour of any of them cannot be fully understood in isolation from the others: ambitious observing programmes like the Canadian Galactic Plane Survey, providing tracers of all of the major ISM components at the same high resolution, are clearly necessary for progress to be made. The physical processes that shape the Milky Way's ISM are not unique to our Galaxy though, they can also be studied in nearby galaxies where the different perspective can make the study of global and large-scale properties much easier, although finer details become obscured due to the loss of resolution. 

There is now an overlap in the scales which are accessible to magnetic field studies of the Milky Way, where the dramatic increase in background rotation measures provided by the CGPS and other surveys is leading to a much clearer understanding of the regular magnetic field structure of the Galaxy (see Brown in this volume), and external galaxies, where high resolution observations can probe magnetic fields down to scales of around 100pc. Over the last two decades the polarized radio emission from several tens of nearby disc galaxies have been observed with a resolution sufficient to reveal the structure of their regular (or mean or large-scale) magnetic field. Thus theoretical models for the origin of galactic magnetic fields can be informed by and tested against observations of more than one system: a useful theory will be at home as equally in the Milky Way as elsewhere. This requires quantifiable properties that can be (relatively) easily derived from the observations and that are also closely linked to the theory. For the mean-field galactic dynamo theory two such properties are the pitch angle of the regular magnetic field and the pattern of azimuthal variation in its strength and direction.

Magnetic fields are present throughout the Universe and appear to be important in many astrophysical processes, such as solar and stellar activity cycles and the formation of jets and accretion discs. Dynamos, at a fundamental level operating in a similar fashion to the galactic dynamo, are thought to play a role in the generation and organisation of magnetic fields in all of these objects. The key ingredients of these dynamos are a source of kinetic energy, generally arising from differential rotation, and a helical component of the turbulent flow in the system. Galaxies are unique among all of these astrophysical objects though, in that these key ingredients are \emph{directly observable} (at least in principle). This, coupled with the availability of readily observable properties that are directly related to the operation of the dynamo, makes the study of G/galactic magnetism one of the most promising avenues for understanding the process of magnetic field generation in the Universe. 

Here I review some of the basic observed properties of galactic magnetic fields, concentrating on two aspects of the regular magnetic field that are closely connected to basic mean field dynamo theory. The selection and discussion of observational and theoretical material is not at all comprehensive: for more on the observational picture, see \citep{Beck:1996,Beck:2005b,Krause:2009} and for discussion of the theory in far greater depth, including many outstanding problems that I have neglected, see \citep{Ruzmaikin:1988,Beck:1996,Widrow:2002,Brandenburg:2005,Shukurov:2007,Kulsrud:2008}.

\section{Basic ideas}
\subsection{Radio continuum observations of nearby galaxies}

Synchrotron radiation at cm wavelengths is produced by the acceleration of GeV cosmic ray electrons moving in the micro-Gauss strength interstellar magnetic field. The power-law distribution of cosmic ray energies results in a power-law distribution of synchrotron intensity $I_\mathrm{syn}$ with frequency $\nu$, with
\begin{equation}
\label{eq:syn}
I_\mathrm{syn}\propto n_\mathrm{cr}\nu^{-q}B_{\perp}^{\,q+1},
\end{equation} 
where $n_\mathrm{cr}$ is the number of cosmic ray electrons (at a particular reference energy), $q$ is the spectral index with typical values in the range $0.5$--$1.0$ and $B_{\perp}$ is the magnetic field in the plane of the sky. So wherever synchrotron radiation is observed we know there are interstellar magnetic fields. 

Galactic synchrotron radiation is inherently polarized, with the observed plane of polarization of the electric field orthogonal to $B_\perp$ in the emitting region. The degree of polarization can theoretically reach a maximum of 70--75\% but in practice this is fraction is reduced due to $B_\perp$ varying in orientation within the telescope beam and due to the superposition of different amounts of Faraday rotation along and across the line of sight through the emitting region and any foreground Faraday screens. Faraday rotation is wavelength ($\lambda$) dependent and is sensitive to the line of sight component of the magnetic field $B_\parallel$, with the observed polarization angle $\psi$ related to the emitted angle $\psi_0$ by, 
\begin{equation}
\label{eq:FD}
\psi=\psi_0(B_\perp)+\mathcal{R}(B_\parallel)\lambda^2=K\int n_\mathrm{e}B_\parallel\mathrm{d}l\ \lambda^2
\end{equation}
where $K$ is a constant whose value depends on the units of the variables, $n_\mathrm{e}$ is the density of thermal electrons and the integral is taken along the line of sight. The quantity $\mathcal{R}$ is often called the Faraday depth. High resolution multi-frequency observations thus, in principle, allow both $B_\perp$ and $B_\parallel$ to be determined as functions of position in a nearby galaxy. In practice the problems of Faraday depolarization \citep{Sokoloff:1998} may be non-trivial to account for, but careful modelling can sometimes reveal otherwise difficult to recover information, such as the relative scale-heights of the relativistic and thermal electrons \citep{Fletcher:2004}.  

The interpretation of the observed radio continuum intensity is complicated by emission from non-relativistic electrons which has nothing to do with magnetic fields; the observed intensity is the sum of both synchrotron and thermal intensities, $I_\mathrm{obs}(\nu)=I_\mathrm{syn}(\nu)+I_\mathrm{th}(\nu)$, where
\begin{equation}
\label{eq:therm}
I_\mathrm{th}\propto\nu^{-0.1},
\end{equation}
at cm wavelengths. The much flatter spectral index of the thermal emission, compared to the synchrotron emission, means that the thermal fraction $f$ becomes higher with increasing frequency. In principle multi-wavelength observations allow the components to be separated, by determining $q$ and the thermal fraction. In a survey of 74 nearby galaxies \citet*{Niklas:1997} found $f=8$\% at $\lambda$30\,cm rising to $f=30$\% at $\lambda$3\,cm, with $q=0.83\pm0.02$. However, these are global results, derived from the integrated emission from the entire galaxy and do not tell us anything about the spatial variation of the thermal emission across an individual galaxy.

It has become standard to assume that $q$ is constant across the disc of a resolved galaxy, whose value can be mildly constrained by examining the variation in the observed spectral index between inter-arm and HII regions, in order to isolate the synchrotron emission. But $q$ is affected by cosmic ray energy losses, due to synchrotron radiation itself and inverse Compton scattering, and so will increase as the electrons travel away from supernova remnants, the sites of their acceleration where $q\simeq 0.5$. A new approach to this problem was recently developed by \citet{Tabatabaei:2007} who used an extinction-corrected H$\alpha$ map of the galaxy M33 to derive the thermal emission measure and hence separate $I_\mathrm{th}$ and $I_\mathrm{syn}$. \citet{Tabatabaei:2007} found $\langle f\rangle=17.6\pm0.4$\% at $\lambda$20\,cm, where $\langle\dots \rangle$ denotes averaging, with $q$ varying between 0.6 in star forming regions to 1.2 in the inter-arms. The old method of assuming a constant $q$ produces a 20\% overestimate of the overall thermal fraction, but more significantly $\langle f\rangle$ masks strong regional variations with HII regions generally having $f>60$\% and the inter-arm regions $f<25$\% at $\lambda$20\,cm. In turn, local variations in $q$ and $f$ mean that the distribution of $I_\mathrm{syn}$ in nearby galaxies derived using an assumed constant $q$, as is common practice, can be significantly different from the actual local synchrotron intensity.

\subsection{Galactic dynamo theory}
\label{sec:dyn}

An astrophysical dynamo is a mechanism for converting the kinetic energy of a flow into magnetic energy: in order for it to work the generation of magnetic field must be faster than its decay due to dissipation into heat. Maxwell's equations are used to derive the magnetic induction equation,
\begin{equation}
\label{eq:ind}
\frac{\partial\mathbf{B}}{\partial t}=\nabla\times(\mathbf{V}\times\mathbf{B})+\eta\nabla^2\mathbf{B},
\end{equation}
where $\eta$ is the magnetic diffusivity. In the simplest models (which are sufficient for the comparison with observations I will make later) the magnetic and velocity fields are split into mean and fluctuating components $\mathbf{B}=\langle\mathbf{B}\rangle+\mathbf{b}$ and $\mathbf{V}=\langle\mathbf{V}\rangle+\mathbf{v}$ and the resulting term $\langle \mathbf{v}\times\mathbf{b}\rangle$, representing the electromotive force arising from the turbulent motion of the magnetic field, is replaced by $\alpha\mathbf{B}+\eta_\mathrm{T}\nabla\times\mathbf{B}$, where $\alpha$ quantifies the helical nature of the turbulent velocity generated by the combined effects of buoyancy and the Coriolis force, and $\eta_\mathrm{T}>\eta$ is the enhanced magnetic diffusivity due to turbulence \citep{Krause:1980,Ruzmaikin:1988,Beck:1996,Widrow:2002}.  

Local models for magnetic field generation can be derived from Equation~(\ref{eq:ind}) by assuming rotational symmetry (no $\phi$ dependence), a thin disc (with the scaleheight $h\ll r$ so that $z$ derivatives are greater than $r$ derivatives) and a regular magnetic field that lies in the galactic plane $\mathbf{B}=(B_r,B_\phi,0)$. We obtain a pair of coupled partial differential equations,
\begin{eqnarray}
\label{eq:Bp}
\frac{\partial B_\phi}{\partial t} & = & r\frac{\mathrm{d}\Omega}{\mathrm{d}r}B_r+\eta_\mathrm{T}\frac{\partial^2 B_\phi}{\partial z^2}, \\
\label{eq:Br}
\frac{\partial B_r}{\partial t} & = &
-\frac{\partial}{\partial z}(\alpha B_\phi)+\eta_\mathrm{T}\frac{\partial^2 B_r}{\partial z^2},
\end{eqnarray}  
where the mean velocity is simple differential rotation $\mathbf{V}=\mathbf{\Omega} r=(0,\Omega r,0)$. Equations~(\ref{eq:Br}) \& (\ref{eq:Bp}) capture the essence of the dynamo process: shear, the first term on the right in Equation~(\ref{eq:Bp}), converts $B_r$ into $B_\phi$ and the $\alpha$-effect, small-scale helical motions with the same sign across a given side of the disc, produce $B_r$ from $B_\phi$. Solutions have the form $B_r,\,B_\phi\propto e^{st}$ and so a dynamo will operate when the real part of $s$ is positive. This condition is met when the dynamo number, the dimensionless combination
\begin{equation}
\label{eq:dyn}
D=\frac{|G\alpha| h^3}{\eta^2_\mathrm{T}},
\end{equation}      
where $G=r \mathrm{d}\Omega/\mathrm{d}r$, exceeds a critical value $D>D_\mathrm{c}\approx 8$ \citep{Ruzmaikin:1988}. Equation~(\ref{eq:dyn}) is readily obtained by replacing z-derivatives with $1/h$ and $1/h^2$ and setting the determinant of the resulting coupled equations to zero. A useful estimate for the magnitude of the $\alpha$-effect is given by 
\begin{equation}
\label{eq:alpha}
|\alpha|\sim \mathrm{min}(\Omega l^2/h,v)
\end{equation}
where $l$ is the correlation scale of the turbulence and $v$ the magnitude of the turbulent velocity \citep{Beck:1996}.

One of the important features of this simple dynamo model is that \emph{both} $B_r$ and $B_\phi$ are generated, so that the resulting regular magnetic field will have a pitch angle given by \citep{Shukurov:2007}
\begin{equation}
\label{eq:pa}
\tan{p_\mathrm{B}}=\frac{B_r}{B_\phi}\approx \sqrt{\frac{|\alpha|}{|G|h}}.
\end{equation} 
On the other hand, amplification of an initial magnetic field solely by the stretching associated with the differential rotation would result in a tightly wound magnetic structure with $p_\mathrm{B}\rightarrow 0\deg$ and which would frequently change sign in both radius and azimuth \citep[see][Sect. 7.5.1 for more details]{Shukurov:2007}.  

Finally, \citet{Ruzmaikin:1988} show that solutions with different azimuthal and vertical  symmetries have different growth rates. In particular the azimuthal mode $m=0$, which describes a regular magnetic field with full rotational symmetry in the $(r,\phi)$ plane, has a greater growth rate than higher modes. Similarly, quadrupolar fields, where $B_r$ and $B_\phi$ are even in $z$ and $B_z$ is odd, are preferred to dipolar fields.

\section{Observed properties of magnetic fields in nearby galaxies}
\subsection{Magnetic field strength}
\begin{figure}
	\begin{center}
	\includegraphics[width=0.8\textwidth]{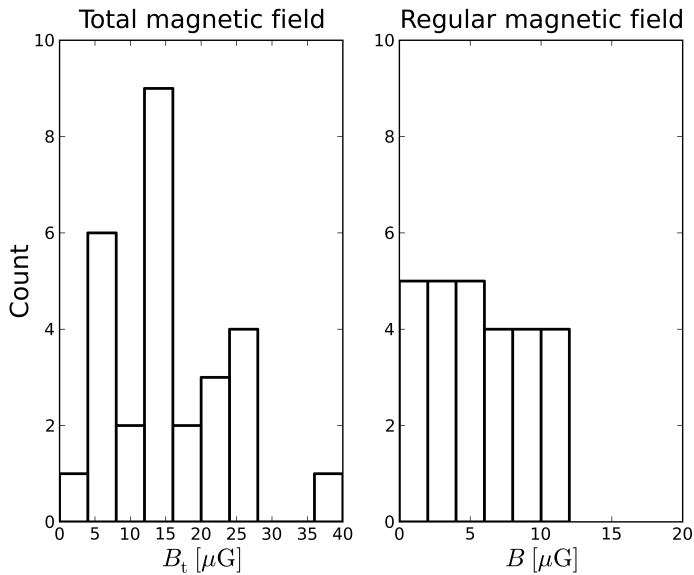}
	\end{center}
	\caption{Distribution of typical values for the total (left) and regular (right) equipartition magnetic field strength for 21 galaxies observed since 2000. Several galaxies have more than one typical field strength quoted (e.g. separate values are given for spiral arms and inter-arm regions) so the total count is greater than the number of galaxies.}
	\label{fig:Bstrength}
\end{figure}

In Equation~(\ref{eq:syn}) strong $I_\mathrm{syn}$ can be due to low-energy cosmic ray electrons interacting with a strong magnetic field or high-energy electrons in a weak field. The most common method used for estimating the strength of the magnetic field from radio observations of synchrotron emission, once the thermal component has been removed, assumes that the energy densities of the magnetic field and cosmic rays are equal. This equipartition method is discussed in detail in \citet{Beck:2005a} and requires information about the energy density of the cosmic ray nucleons relative to the electrons (most of the energy lies with the heavier particles but these do not contribute to $I_\mathrm{syn}$), the size of the emitting region and the filling factor of the magnetic field (the latter quantity, quantifying the intermittent nature of the field, is a potentially important diagnostic of the field origin about which very little is currently known). An estimate for the strength of the regular magnetic field $B$ can be made from the degree of polarization of the synchrotron emission. 

Fortunately the derived equipartition magnetic field strengths turn out to depend only weakly on these input parameters. Unfortunately there is no compelling astrophysical reason why equipartition should hold, particularly on scales of 1kpc or less. A few tests have been made on scales $>1$kpc though. \citet{Hummel:1986} compared the integrated radio and infra-red fluxes of 65 Sbc galaxies and concluded that globally energy equipartition held for most of the galaxies. \citet*{Strong:2000} used a model for the azimuthally averaged $B(r)$, where $r$ is the radius, for the Milky Way that is compatible with $\gamma$-ray data and which agrees well with the equipartition values derived by Berkhuijsen \citep[in][]{Beck:1996}. Where estimates for the regular magnetic field strength have been made using Faraday rotation data these are often in good agreement with equipartition estimates \citep[e.g.][]{Fletcher:2004,Beck:2007}, and although this is not always the case \citep[e.g.][]{Beck:2005,Fletcher:2010} these differences seem to arise when a significant fraction of the ordered magnetic field is due to an anisotropic random component, which results in strong polarized radio emission but does not contribute much Faraday rotation. 

Using the integrated radio continuum emission \citet{Hummel:1986} found $\langle B_\mathrm{tot} \rangle = 8\mu$G for 65 Sbc galaxies, \citet{Fitt:1993} obtained $\langle B_\mathrm{tot} \rangle = 10\mu$G (using 100 for the ratio of electron to nucleon energies rather than the value of 0 adopted in their paper) for 146 late-type galaxies and \citet{Niklas:1995} derived $\langle B_\mathrm{tot} \rangle = 9\mu$G for 74 nearby galaxies. Uncertainties due to the adopted parameter values probably amount to around 30\% \citep{Beck:1996}. All of these results give the average total magnetic field strength based on the integrated radio emission of the galaxies; since these results were published the increasing number of high resolution observations of nearby galaxies means that it has become common to derive equipartition magnetic field strengths for particular regions within a galaxy. It is worth reiterating that indications that equipartition is a valid assumption have been derived from the average properties of galaxies on large-scales. 

Figure~\ref{fig:Bstrength} shows the distributions of ``typical" values of the total and regular equipartition magnetic field strengths for 21 galaxies drawn from the recent (post-2000) literature. Several papers identify separate typical field strengths for, for example, spiral arms and inter-arm regions and I have included these as separate data points in Figure~\ref{fig:Bstrength}. Thus, while these histograms do not show distributions of $\langle B_\mathrm{tot} \rangle$ and $\langle B \rangle$ that are directly comparable to the well defined samples of \citet{Hummel:1986,Fitt:1993,Niklas:1995} they do illustrate recent estimates for the strength of ``typical" interstellar magnetic fields for a group of spiral, barred, starburst, dwarf and irregular galaxies. The mean total field in this data is $B_\mathrm{tot}=17\mu\mathrm{G}$, with a standard deviation of $14\mu\mathrm{G}$, a factor of two higher than the average field strengths quoted above. For the regular field the mean is $B=5\mu\mathrm{G}$, with standard deviation $3\mu\mathrm{G}$; thus the random component of the field is around 3 times the strength of the regular component. 

Regular magnetic fields of these magnitudes can be generated in a galactic lifetime by the dynamo if a seed field of cosmological origin ($B_\mathrm{tot}<10^{-20}$G) is first amplified by a small-scale (or fluctuation) dynamo: these dynamos, that occur in random turbulent flows, have a faster growth rate than the mean field dynamo. The mean field dynamo is still required, however, in order to produce the galaxy-scale regular magnetic field component \citep[see][Sect. 5.3]{Beck:1996}.

\subsection{Magnetic field patterns}
\begin{figure}
	\begin{center}
	\includegraphics[width=0.48\textwidth]{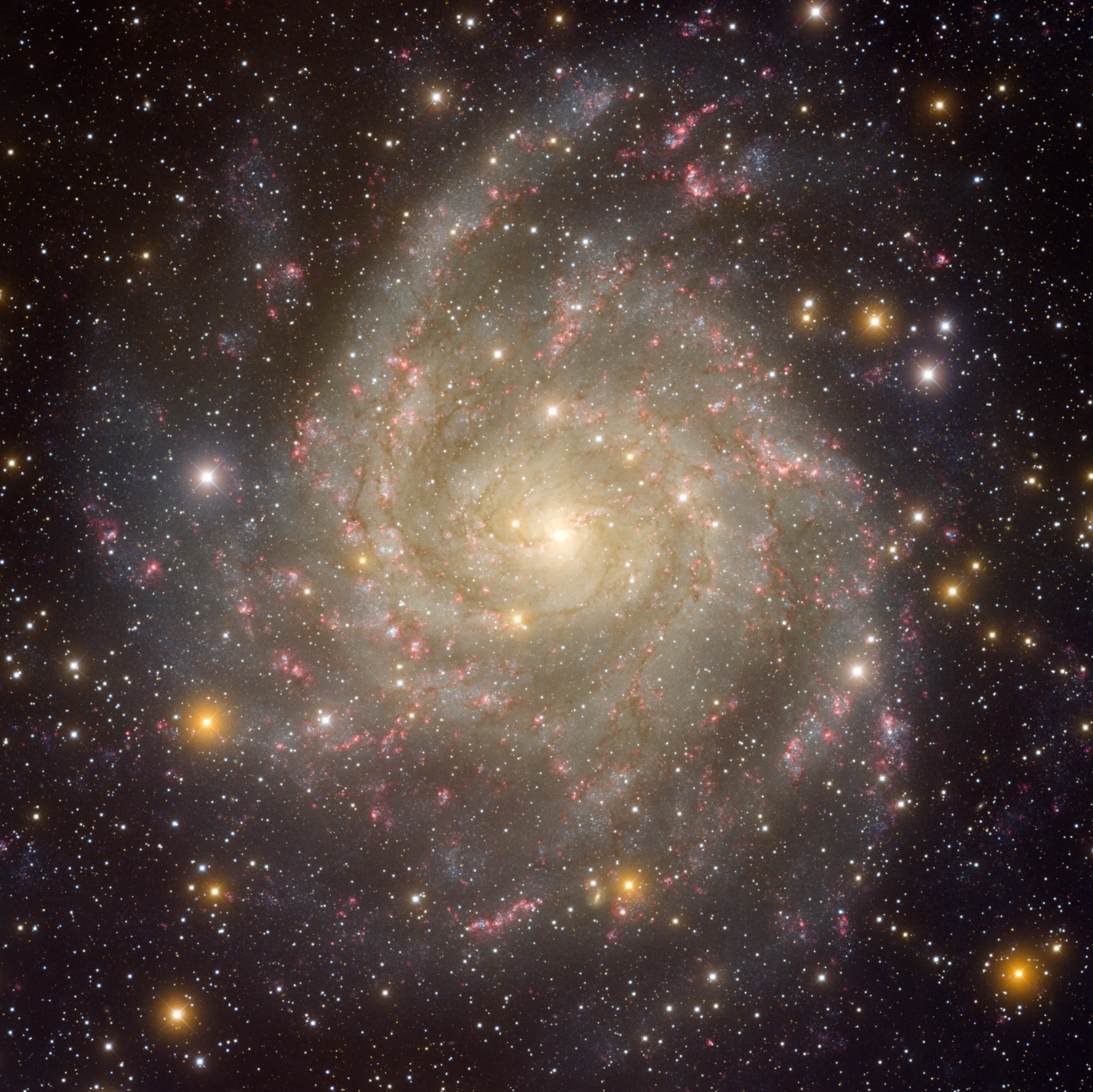}\hfill
	\includegraphics[width=0.48\textwidth]{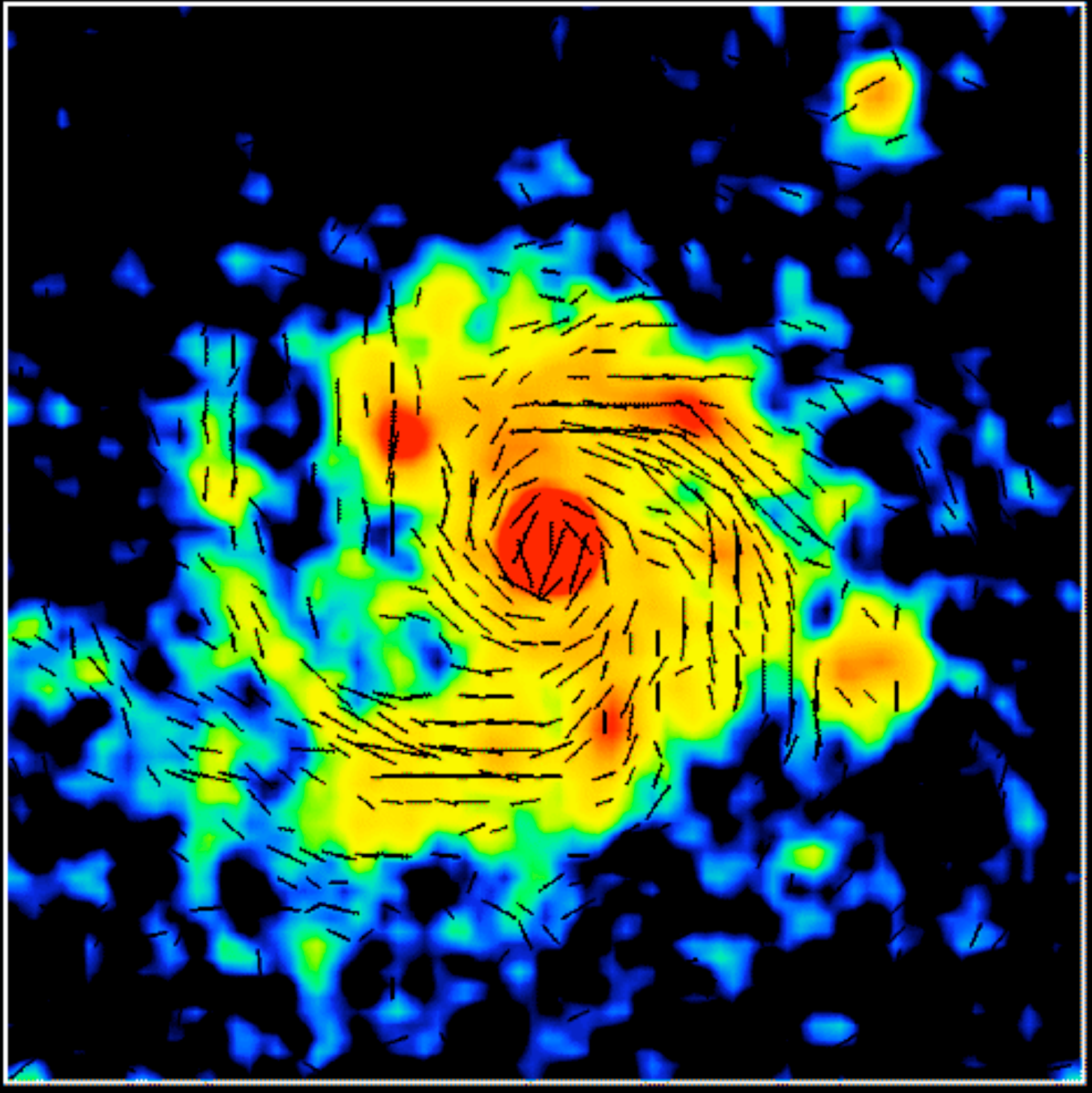}
	\includegraphics[width=0.48\textwidth]{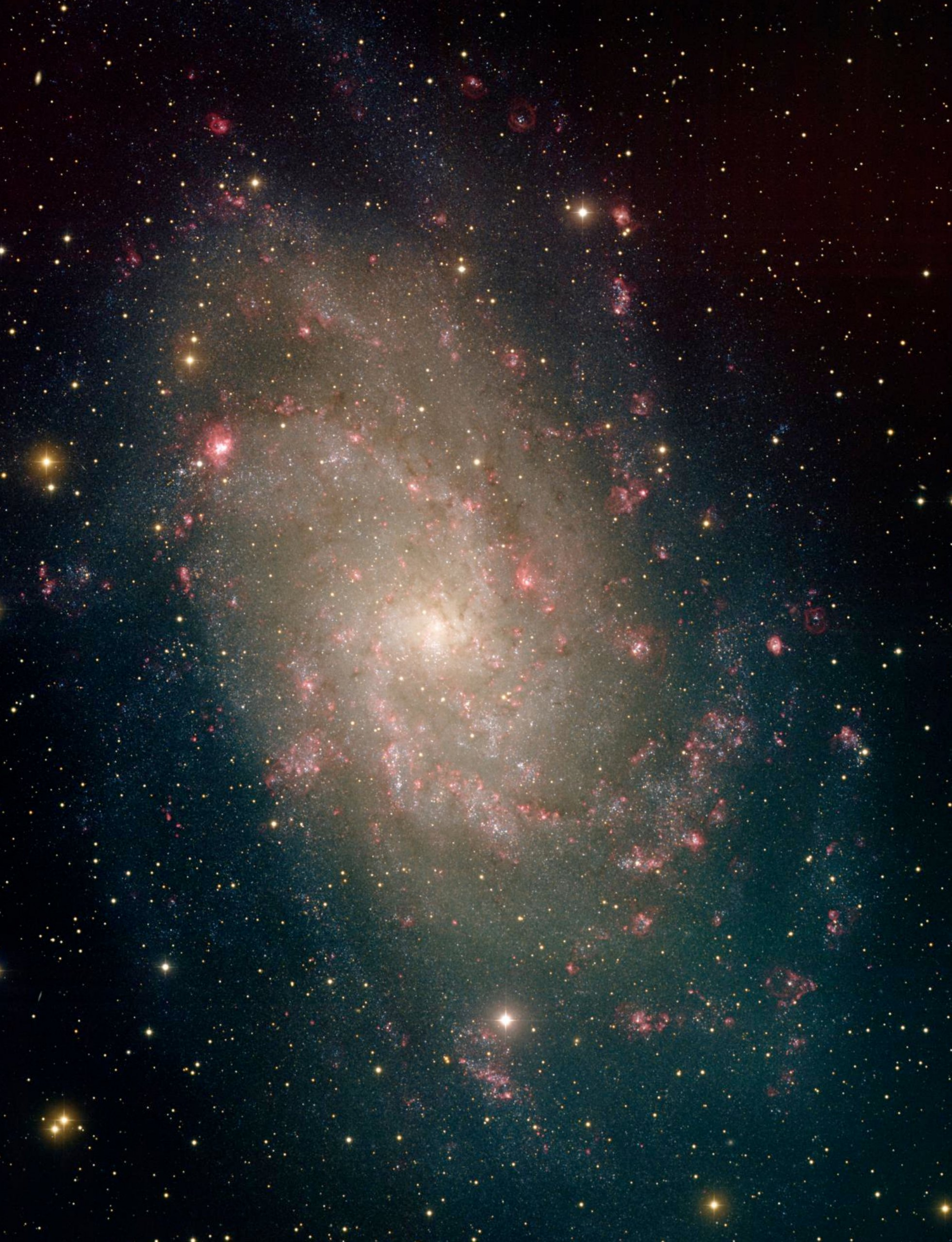}\hfill
	\includegraphics[width=0.48\textwidth]{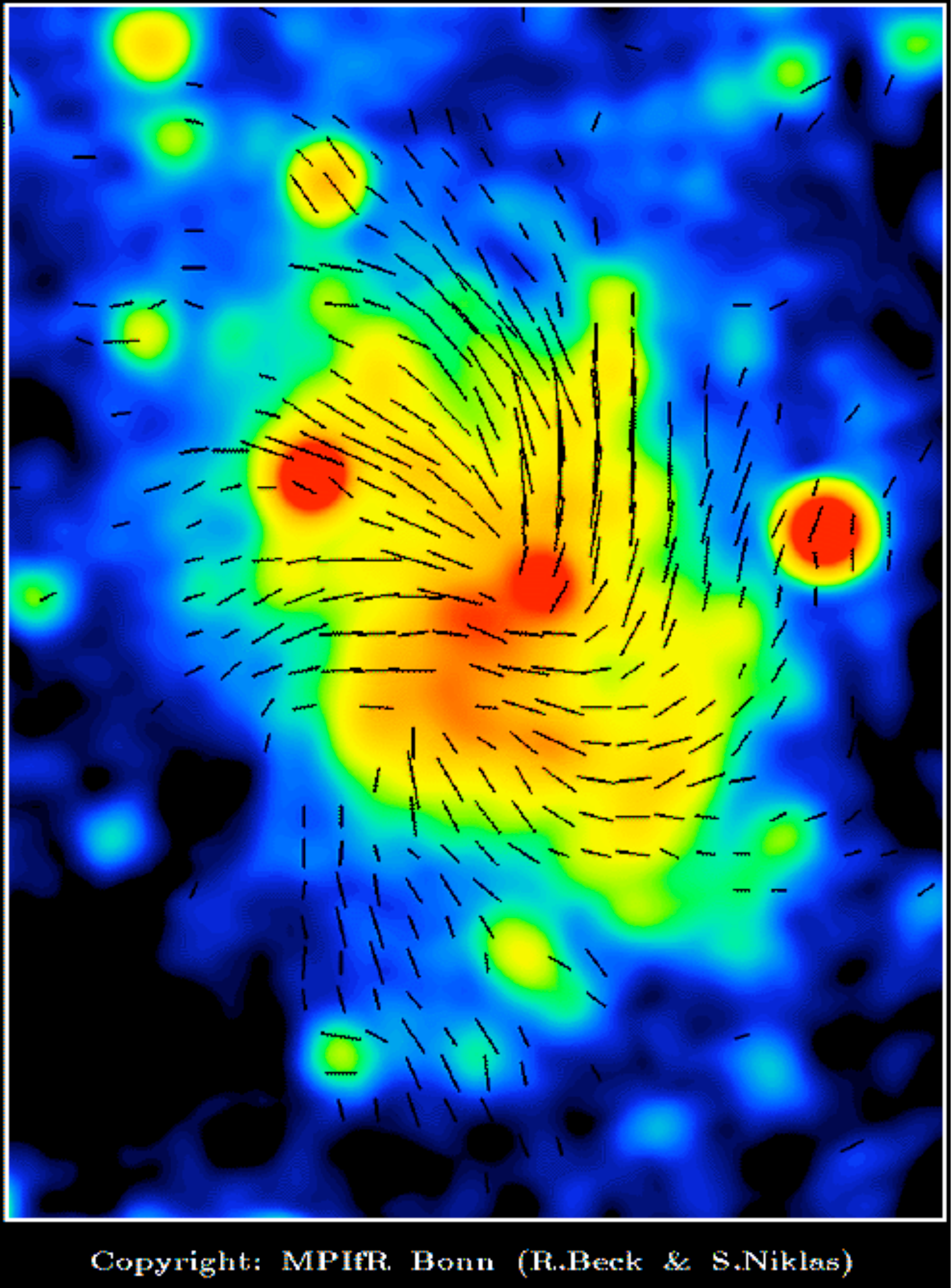}
	\end{center}
	\caption{\textbf{Upper row:} the spiral galaxy IC342. On the left the well defined spiral arms are shown in an optical image. The right panel shows Effelsberg observations of the $\lambda$3cm total radio emission and the orientation of the polarization B-vectors, which follow a spiral pattern.  \textbf{Lower row:} the flocculent galaxy M33. On the left spiral arms are less well defined than those of IC 342 in an optical image. In the right panel, Effelsberg $\lambda$6\,cm total intensity and B-vectors show that the magnetic field also follows a large-scale spiral pattern. Radio maps are from the MPIfR `Atlas of magnetic fields in nearby galaxies' http://www.mpifr-bonn.mpg.de/div/konti/mag-fields.html copyright and courtesy of R. Beck and the MPIfR, Bonn. Optical image credits: IC342, T.A. Rector/University of Alaska Anchorage, H. Schweiker/WIYN and NOAO/AURA/NSF; M33, T.A. Rector (NRAO/AUI/NSF and NOAO/AURA/NSF.}	
	\label{fig:spirals}
\end{figure}

\begin{figure}
	\begin{center}
	\includegraphics[width=0.95\textwidth]{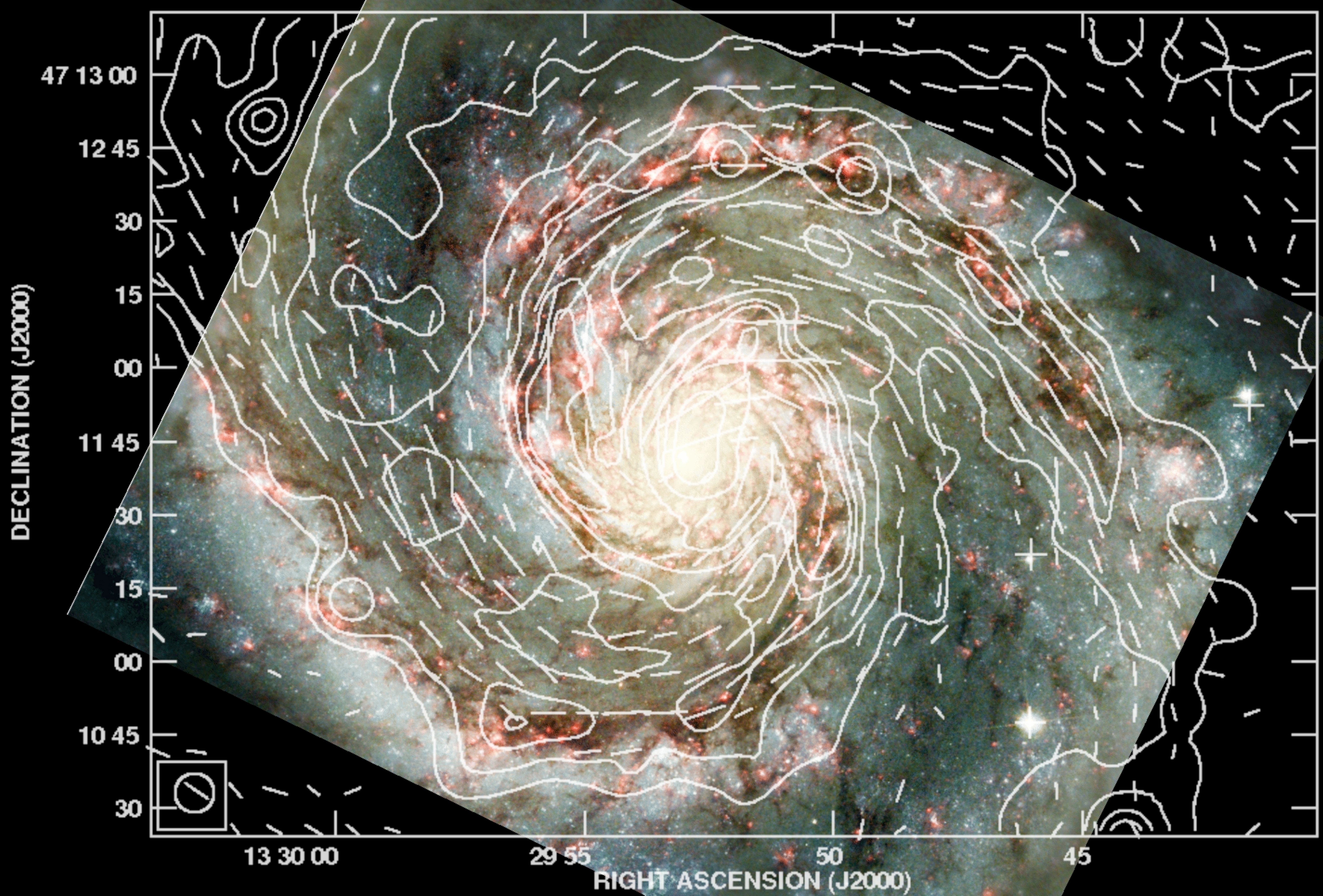}
	\end{center}
	\caption{The $\lambda$6\,cm total radio emission (VLA and Effelsberg combined) in the
central $3\arcmin \times 4\arcmin$ of M51 at $8\arcsec$ resolution, corresponding to 300\,pc at the distance of M51, overlaid on a Hubble
Space Telescope image ({http://heritage.stsci.edu/2001/10}, image credit: NASA
and The Hubble Heritage Team (STScI/AURA)). Contours are at $6, 12, 24, 36, 48,
96, 192$ times the noise level of $25\mu\,Jyb$. Faraday rotation corrected
$B$-vectors, showing the orientation of the regular magnetic field, are plotted
where the polarized intensity $P\ge 3\sigma_{P}$. The radio data is from \citet{Fletcher:2010}.}
	\label{fig:m51}
\end{figure}

A large-scale magnetic field component has been observed in most classes of galaxies (the exception being ellipticals)\footnote{Only a single representative reference is given in each case: these have been selected in an attempt to cover examples from all groups active in the field.}: spirals \citep{Braun:2010}, cluster spirals \citep{Chyzy:2008a}, barred \citep{Beck:2002}, flocculent \citep{Tabatabaei:2008}, dwarf \citep{Mao:2008}, irregular \citep{Gaensler:2005}, starburst \citep{Kepley:2010}. In all cases where the galaxy is sufficiently face-on to allow the disc-plane magnetic pattern to be determined the large-scale fields clearly trace out a spiral pattern (Figure~\ref{fig:spirals}), (the lone exception being the SMC \citep{Mao:2008} whose large-scale field is uni-directional, probably due to the influence of the Magellanic Bridge). Particularly striking are the magnetic field lines that: trace a continuous spiral from the nucleus to the outermost regions of M51 (Figure~\ref{fig:m51}) \citep{Neininger:1992,Fletcher:2010}; that cut across, and are apparently unperturbed by, the well defined ring of active star formation in M94 \citep{Chyzy:2008b} and that form coherent galaxy-scale spiral structures in the irregular, weakly rotating galaxy NGC~4449 \citep{Chyzy:2000}. 

It is necessary to be careful regarding terminology here. The observed polarization angles, when corrected for Faraday rotation (which is generally necessary for wavelengths in excess of $\lambda$6cm) and rotated by 90$\deg$, to produce what are often called B-vectors, indicate the presence of an ordered magnetic field component on scales of the telescope beam and greater. This ordered large-scale field is not necessarily a mean field in the sense of a mean component of the magnetic vector $\mathbf{B}$ \citep{Han:1999}: it is not difficult to imagine the action of shear and compression producing anisotropy, and thus order, in an initially purely isotropic random magnetic field. Such strong anisotropic large-scale magnetic fields have been identified in M51 \citep{Fletcher:2010} and the barred galaxies NGC~1097 and NGC~1365 \citep{Beck:2005}. If the large-scale magnetic field shows coherence in its Faraday rotation, which is sensitive to the sign of the magnetic field, then we may refer to it as a mean or regular magnetic field component.

\subsubsection{Pitch angles}

In Section~\ref{sec:dyn} we saw that a mean field dynamo will generate both radial and azimuthal components of the regular magnetic field. \emph{All} galaxies that have been observed with suitable resolution are found to contain a large-scale component of their magnetic field with a non-zero pitch angle.  

In the more face-on spiral galaxies it often appears that the pitch angle of the magnetic field is the same as the optical or gaseous spiral arms (e.g. Figure~\ref{fig:m51}). Table~\ref{tab:pa} shows the pitch angles of the regular magnetic field $p_{B}=\mathrm{arctan}(B_r/B_\phi)$, recovered from multi-wavelength observations using models that take into account the Faraday rotation. These have been split, where possible, into values for the inner and outer galaxy. In addition pitch angles for the optical or gaseous spiral arms $p_\mathrm{sp}$, drawn uncritically from the literature, are given: these tend to have been be obtained by fitting logarithmic spirals to images. This Table is the first attempt at trying to systematically quantify the degree of correspondence between well defined regular magnetic field pitch angles and spiral arm pitch angles in a (small) sample of galaxies. 

\begin{table}[!ht] 
\caption{Pitch angles of the magnetic field and spiral arms}
\label{tab:pa}
\smallskip
\begin{center}
{\small
\begin{tabular}{lcccl}
\noalign{\smallskip}
Galaxy & \multicolumn{2}{c}{Regular magnetic field} & Spiral arms & Reference \\
\noalign{\smallskip}
\cline{2-3}
\noalign{\smallskip}
       & inner & outer & & \\
\noalign{\smallskip}
\tableline
\noalign{\smallskip}
IC 342  & $-20\pm2\deg$ & $-16\pm2\deg$ & $-19\pm5\deg$ & \citet*{Krause:1989a} \\
M31     & $-17\pm4\deg$ & $-8\pm3\deg$ & $-7\deg$ & \citet{Fletcher:2004} \\
M33     & $-48\pm12\deg$ & $-42\pm5\deg$ & $-65\pm5\deg$ & \citet{Tabatabaei:2008} \\
M51     & $-20\pm1\deg$ & $-18\pm1\deg$ & $-20\deg$ & \citet{Fletcher:2010} \\
M81     & $-14\pm7\deg$ & $-22\pm5\deg$ & $-11\deg\rightarrow -17\deg\ ^{a}$ & \citet*{Krause:1989b} \\
NGC6946 & \multicolumn{2}{c}{$-38\pm12\deg\ ^{b}$} &  $-36\pm7\deg\ ^{b}$ & \citet{Frick:2000} \\
\tableline
\end{tabular}
}
\end{center}
Negative pitch angles denote trailing spirals.\\
$^{a}$ Change of pitch angle from inner to outer galaxy.\\
$^{b}$ These are average values for 5 spiral arms that have different radial extents.
\end{table}

Table~\ref{tab:pa} shows that the magnetic and optical spirals do have the same pitch angles in five of the six galaxies, at least across a significant part of the galaxy's disc. The only galaxy where the spiral arms and regular magnetic field have significantly different pitch angles is the flocculent galaxy M33, shown in the lower panels of Figure~\ref{fig:spirals}. M33 contains many short segments of spiral arm but lacks the "grand design", galaxy scale, spiral arms seen in the other galaxies of Table~\ref{tab:pa}. 

The close alignment between spiral arms and regular magnetic fields, both visually and on average as shown in Table~\ref{tab:pa}, can hide a strong azimuthal variation in the regular magnetic pitch angle: clear examples of this variation can be seen for NGC~6946 in \citet[Fig.~14]{Ehle:1993} and for NGC~4254 in \citet[Fig.~6]{Chyzy:2008a}. \citet{Patrikeev:2006} determined the spiral arm pitch angles in M51 using an anisotropic wavelet decomposition, thus allowing the local pitch angle to be determined rather than the globally best fitting logarithmic spiral, and compared these to the azimuthal variation of $p_{B}$. They found that $p_{B}=p_\mathrm{sp}$ at the position of the spiral arms but varies by $\pm 15\deg$ in the inter-arm regions; these variations are unsystematic whereas \citet{Ehle:1993} found that in NGC~6946 $p_{B}$ is systematically smaller in the inter-arms. The alignment of $p_{B}$ with the spiral arms at the position of the arms is what we would expect if large-scale spiral shocks are present, for example due to density waves in the gas, as the shocks would amplify the component of $B$ parallel to the shock front. The variation of $p_{B}$ away from the arms is not understood: non-circular motions in the velocity field or the dynamo generation of $B$ with $p_{B}\ne p_\mathrm{sp}$ in the inter-arm are two possibilities.

The dynamo theory discussed in Section~\ref{sec:dyn} can be used to derive rough estimates for $p_{B}$ \citep{Shukurov:2007}. Using Equations~(\ref{eq:pa}) \& (\ref{eq:alpha}) we obtain $p_{B}\sim l/h$ for a flat rotation curve. If the properties of the turbulence are roughly constant in radius then gaseous disc flaring with increasing radius means that we should expect $p_{B}$ to decrease with radius. Just this behaviour is shown by four out of five galaxies in Table~\ref{tab:pa}. The exception is M81, where the optical spiral arm pitch angles also increase with radius: we will come back to M81 as an exception in the following Section. 

\begin{figure}
\plottwo{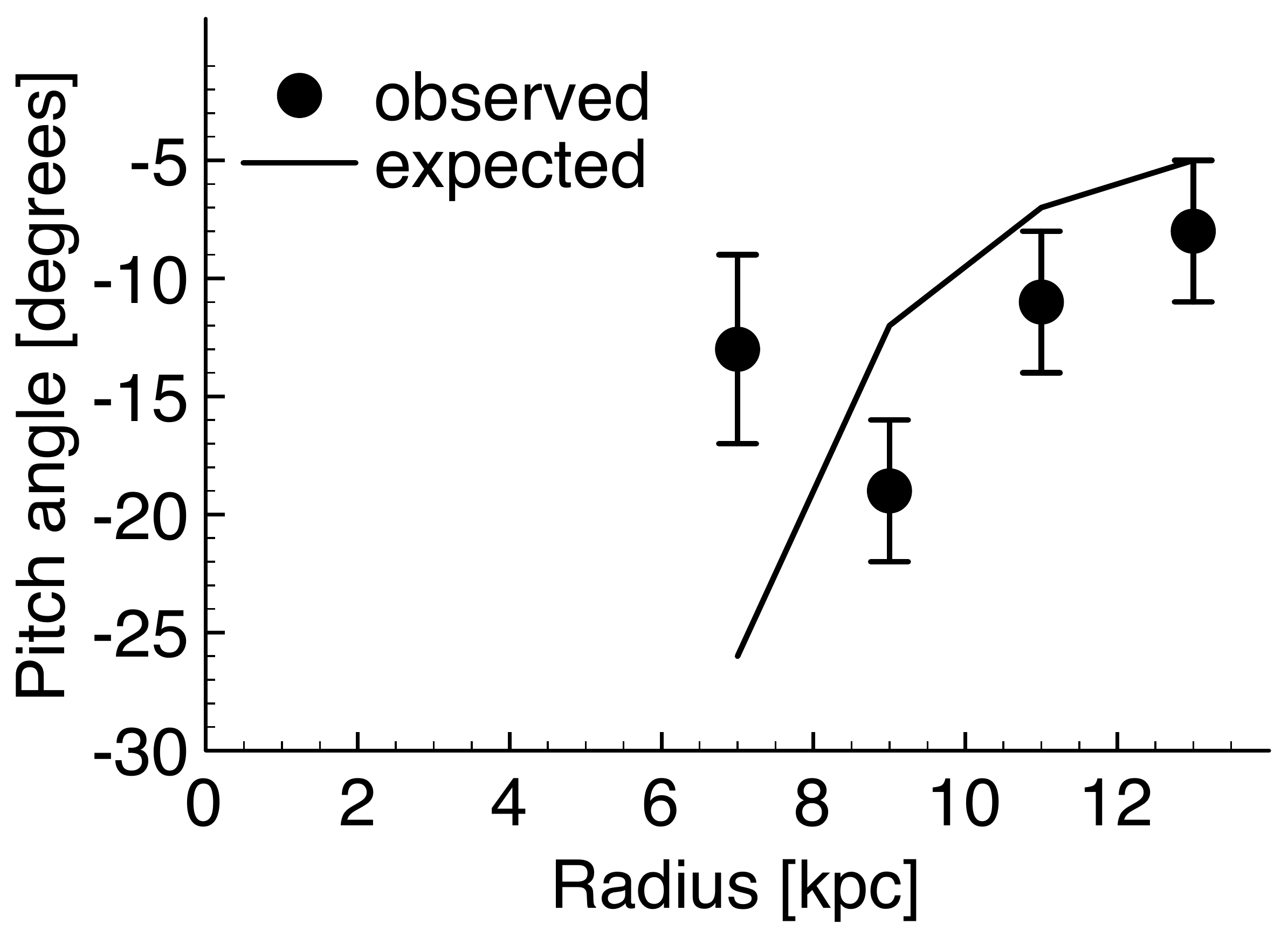}{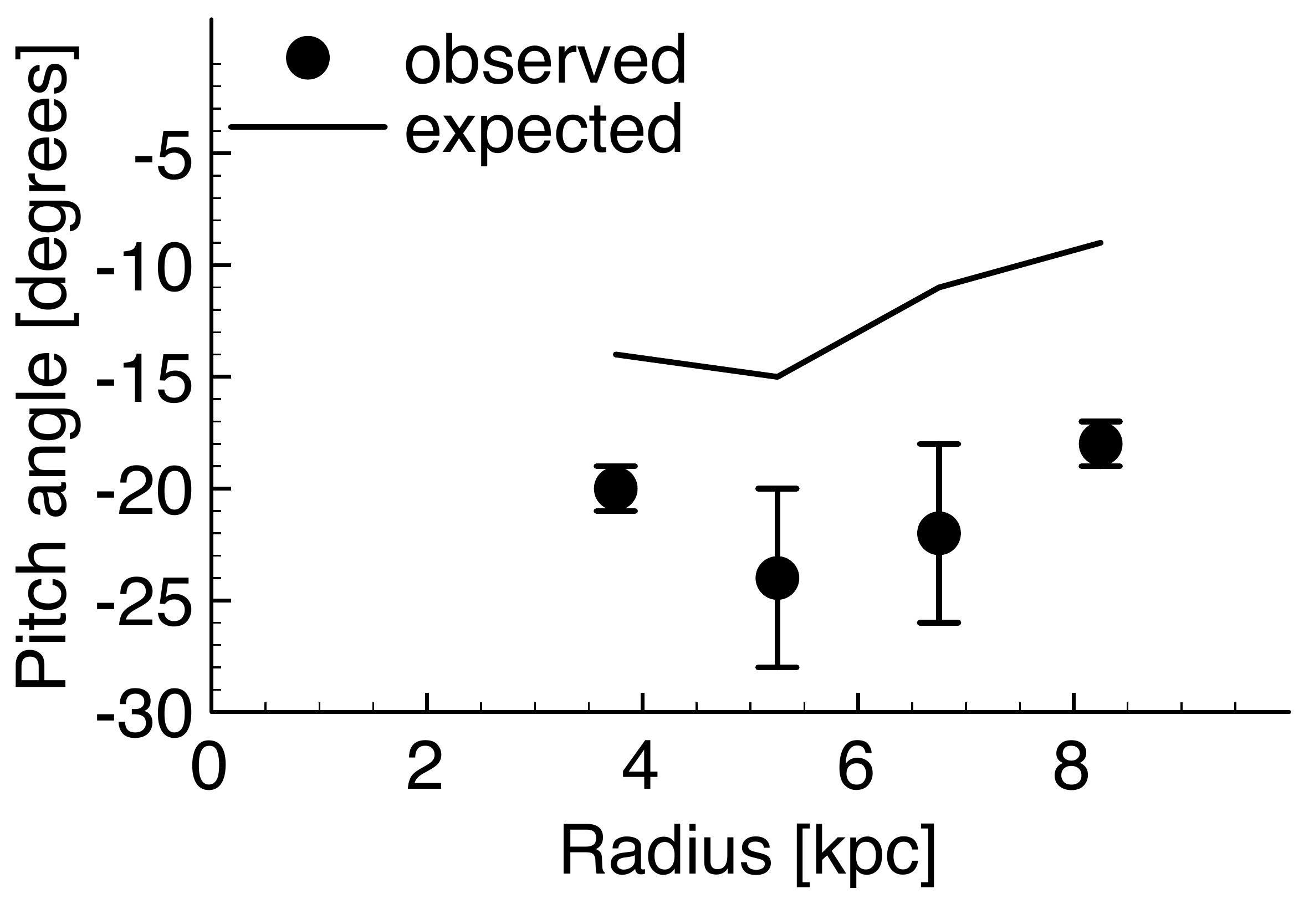}
\caption{Pitch angles of the axisymmetric ($m=0$) regular magnetic field component  derived from observations (points with error bars) and those calculated using Eq.~(\ref{eq:pa}) (lines). \textbf{Left:} M31, using the rotation curve of \citet*{Chemin:2009} and the estimated thermal disc scale-height $h$ and observed $p_\mathrm{B}$ from \citet{Fletcher:2004}. \textbf{Right:} M51, using the rotation curve of \citet*{Garcia-Burillo:1993}, estimated $h$ from \citet{Berkhuijsen:1997} and observed $p_\mathrm{B}$ from \citep{Fletcher:2010}; in both galaxies $l=100$pc was assumed for the correlation scale of the turbulence.}
\label{fig:pa}
\end{figure}

Equation~(\ref{eq:pa}) can be used to estimate $p_{B}\simeq -$(10--20)$\deg$, given a typical flat rotation curve, $h\simeq 1$kpc for the scaleheight of the magneto-ionic disc  and $\alpha\simeq 1$km~s$^{-1}$. Figure~\ref{fig:pa} shows a comparison of the observed $p_{B}$ for the $m=0$ azimuthal mode (possibly produced by a dynamo, see below) with that predicted by Equation~(\ref{eq:pa}) for two nearby galaxies, using observed rotation curves and estimates for disc flaring. While the agreement between expected and observed pitch angles is not exact, the magnitudes are similar and there is a close correspondence in the radial variation, which is remarkable given the small number of parameters involved and the simplicity of the model. More detailed dynamo models created specifically for these two galaxies will surely produce a closer agreement: indeed the spatial variation of pitch angles is an excellent diagnostic with which to constrain detailed dynamo models for specific galaxies.   

\subsubsection{Azimuthal modes}

The azimuthal structure of the regular magnetic field in the disc of a galaxy can be deduced from observations. Several methods have been developed to do this: examining the azimuthal pattern of the Faraday rotation of diffuse emission \citep{Krause:1989b,Krause:1989a} or that of sources behind the target galaxy \citep{Han:1998,Gaensler:2005}; modelling the pattern of multi-wavelength polarization angles directly \citep{Berkhuijsen:1997,Fletcher:2004}; comparing the azimuthal variations of both Faraday depth and polarized intensity with reference models \citep*{Braun:2010}. 

\begin{table}[!ht] 
\caption{Regular magnetic field azimuthal-mode amplitudes}
\label{tab:mode}
\smallskip
\begin{center}
{\small
\begin{tabular}{lcccl}
\noalign{\smallskip}
Galaxy & \multicolumn{3}{c}{Relative mode amplitude} & Reference \\
\noalign{\smallskip}
\cline{2-4}
\noalign{\smallskip}
       & $m=0$ & $m=1$ & $m=2$ & \\
\noalign{\smallskip}
\tableline
IC342     & 1 & -	& -		& \citet{Krause:1989a} \\ 
LMC       & 1 & -	& -		& \citet{Gaensler:2005} \\
M31       & 1 & 0	& 0		& \citet{Fletcher:2004} \\
M33       & 1 & 1	& 0.5	& \citet{Tabatabaei:2008} \\
M51       & 1 & 0	& 0.5	& \citet{Fletcher:2010} \\
M81       & - & 1	& -		& \citet{Krause:1989b} \\
NGC 253   & 1 & -	& -		& \citet{Heesen:2009} \\
NGC 1097  & 1 & 1	& 1		& \citet{Beck:2005} \\
NGC 1365  & 1 & 1 	& 1		& \citet{Beck:2005} \\
NGC 4254  & 1 & 0.5 & -		& \citet{Chyzy:2008a} \\	
NGC 4414  & 1 & 0.5	& 0.5	& \citet{Soida:2002} \\
NGC 6946  & 1 & -	& -		& \citet{Ehle:1993} \\
\tableline
\end{tabular}
}
\end{center}
Note: a dash indicates that combinations of modes including this entry were not sought whereas a zero indicates that the mode is not present. 
\end{table}

Table~\ref{tab:mode} summarises the approximate amplitudes of the azimuthal modes that have been identified in 12 nearby galaxies: only galaxies where some form of modelling --- ranging from the basic fitting of a sine function to a plot of Faraday rotation, to more sophisticated searches for the best fitting values for parameterized models --- has been used to identify the modes specifically for that galaxy have been included. In many cases I have crudely averaged radial variations. Not included in Table~\ref{tab:mode} are 12 galaxies (two of which are present in the Table) observed by \citep*{Braun:2010}, applying the Rotation measure synthesis method \citep{Brentjens:2005}, which removes some of the complications arising from Faraday depolarization, to external galaxies for the first time. In \citet{Braun:2010} the azimuthal variation of Faraday depth and polarized emission and the varying degree of inclination of the galaxies in the sample was compared to a set of illustrative models, involving different magnetic field topologies, to show that the observed patterns are compatible with a predominantly axisymmetric disc magnetic field.     

Remarkably, the basic, axially symmetric $m=0$ mode is not only present but is the (sometimes equally) strongest mode in all but one of these galaxies, including the irregular LMC, the flocculents M33 and NGC~4414 and the barred NGC~1097 and NGC~1365. If we include the 10 extra galaxies from \citep{Braun:2010}, 21 out of 22 galaxies contain a dominant $m=0$ mode in their disc regular magnetic field. The sole exception is M81, which also stands out among the pitch angles shown in Table~\ref{tab:pa}. Note that the data used in the analysis of M81 are old and not as complete as those available for other galaxies in the list: new data for M81 have recently been obtained and it will be interesting to see if the dominant $m=1$ mode is still found once they are analysed.

The near ubiquitous presence of a strong $m=0$ azimuthal mode is clear evidence in support of the dynamo origin of galactic regular magnetic fields, as this mode is the easiest to sustain under typical conditions. The higher modes that are also found in most galaxies may be due to dynamo action, if the dynamo number is high enough for these modes to have a positive growth rate, or may be due to perturbations of an underlying $m=0$ pattern by dynamical effects of the velocity field due to bars (note that the $m=1$ mode is present in the two barred galaxies, as might be expected) or spiral arms (M51 has a strong two armed spiral pattern and its regular magnetic field contains an $m=2$ mode).  

\subsubsection{Vertical symmetry}

Less is known about the vertical symmetry, or parity, of regular magnetic fields. In principle edge-on galaxies should allow the identification of the vertical structure of the field, but in practice strong depolarization in the disc and a weak line of sight component of the field in the halo make this difficult. The current state of observations is summarised in \citet{Krause:2009}. Recently \citet{Heesen:2009} modelled multi-wavelength observations of NGC~253 and found that a quadrupolar (even parity) disc and halo field is preferred but that a dipolar component of the field in the halo cannot be ruled out. \citet{Braun:2010} found that reference models with quadrupole symmetry best match their sample of 12 galaxies of varying inclination; however the broad classes of azimuthal and vertical symmetries include many different specific field geometries that can be superposed, so fitting of specific parameterized models to this data could produce the best evidence yet for a preferred parity. 

So, the information currently available on vertical parity of the regular magnetic field seems to favour quadrupole symmetry: this is the mode that has the highest growth rate in the mean field dynamo theory \citep{Ruzmaikin:1988,Beck:1996}.  

\section{Summary and future prospects}

Remarkably, the simplest form of the galactic mean field dynamo theory is compatible with, and provides an explanation for, the observed magnetic field pitch angles and azimuthal symmetries that have been accurately determined for a small sample of nearby galaxies. The next steps in both theory and observations seem clear: to create dynamo models for specific galaxies that take as their inputs as many relevant observed properties of the galaxy as possible --- such as rotation curves, non-circular velocities, outflows and star formation rate, gas distribution --- and produce as their output observable properties of the field; to increase the list of galaxies for which observable quantities that can be related to the theory are known, especially through the systematic observation of well chosen samples.    

Both of these tasks are achievable in the near future. As the next generation of radio telescopes such as LOFAR, ASKAP and MeerKAT allow for the routine implementation of new observational methods, such as rotation measure synthesis \citep{Brentjens:2005,Braun:2010} and rotation measure grids \citep*{Gaensler:2004}, which can be interpreted using new techniques \citep{Stepanov:2008}, the database of galaxies with known magnetic field configurations should increase rapidly. Dynamo models for individual galaxies have already been investigated \citep[e.g.][]{Krasheninnikova:1989,Poezd:1993,Moss:1998,Rohde:1999,Moss:2007} and can be readily extended and developed to take into account recent theoretical developments, such as the recognition that transport of magnetic fields out of the disc by galactic winds or fountain flows can be crucial in allowing the dynamo to saturate close to energy equipartition with the ISM turbulence \citep{Shukurov:2006,Sur:2007}. In addition, direct numerical simulations of magnetic field amplification in local \citep{Gressel:2008} and global \citep*{Gissinger:2009} models of the ISM will lead to refinement of the galactic dynamo theory.

\acknowledgements I thank the Royal Society for a travel grant and the organisers for their invitation to a stimulating meeting in a beautiful part of Canada.

\bibliography{fletcher}

\begin{thebibliography}{}
\expandafter\ifx\csname natexlab\endcsname\relax\def\natexlab#1{#1}\fi
\expandafter\ifx\csname url\endcsname\relax
  \def\url#1{\texttt{#1}}\fi
\expandafter\ifx\csname urlprefix\endcsname\relax\def\urlprefix{URL }\fi
\providecommand{\eprint}[2][]{\url{#2}}

\bibitem[{Beck(2005)}]{Beck:2005b}
Beck, R. 2005, in Cosmic Magnetic Fields, edited by R.~Wielebinski, \& R.~Beck
  (Berlin Heidelberg: Springer), vol. 664 of Lecture Notes in Physics, 41

\bibitem[{Beck(2007)}]{Beck:2007}
--- 2007, A\&A, 470, 539

\bibitem[{Beck et~al.(1996)Beck, Brandenburg, Moss, Shukurov, \&
  Sokoloff}]{Beck:1996}
Beck, R., Brandenburg, A., Moss, D., Shukurov, A., \& Sokoloff, D. 1996,
  ARA\&A, 34, 155

\bibitem[{Beck et~al.(2005)Beck, Fletcher, Shukurov, Snodin, Sokoloff, Ehle,
  Moss, \& Shoutenkov}]{Beck:2005}
Beck, R., Fletcher, A., Shukurov, A., Snodin, A., Sokoloff, D.~D., Ehle, M.,
  Moss, D., \& Shoutenkov, V. 2005, A\&A, 444, 739

\bibitem[{Beck \& Krause(2005)}]{Beck:2005a}
Beck, R., \& Krause, M. 2005, AN, 326, 414

\bibitem[{Beck et~al.(2002)Beck, Shoutenkov, Ehle, Harnett, Haynes, Shukurov,
  Sokoloff, \& Thierbach}]{Beck:2002}
Beck, R., Shoutenkov, V., Ehle, M., Harnett, J.~I., Haynes, R.~F., Shukurov,
  A., Sokoloff, D.~D., \& Thierbach, M. 2002, A\&A, 391, 83

\bibitem[{Berkhuijsen et~al.(1997)Berkhuijsen, Horellou, Krause, Neininger,
  Poezd, Shukurov, \& Sokoloff}]{Berkhuijsen:1997}
Berkhuijsen, E.~M., Horellou, C., Krause, M., Neininger, N., Poezd, A.~D.,
  Shukurov, A., \& Sokoloff, D.~D. 1997, A\&A, 318, 700

\bibitem[{Boulares \& Cox(1990)}]{Boulares:1990}
Boulares, A., \& Cox, D.~P. 1990, ApJ, 365, 544

\bibitem[{Brandenburg \& Subramanian(2005)}]{Brandenburg:2005}
Brandenburg, A., \& Subramanian, K. 2005, Physics Reports, 417, 1

\bibitem[{Braun et~al.(2010)Braun, Heald, \& Beck}]{Braun:2010}
Braun, R., Heald, G., \& Beck, R. 2010, A\&A, 514, A42

\bibitem[{Brentjens \& De~Bruyn(2005)}]{Brentjens:2005}
Brentjens, M.~A., \& De~Bruyn, A.~G. 2005, A{\&}A, 441, 1217

\bibitem[{Chemin et~al.(2009)Chemin, Carignan, \& Foster}]{Chemin:2009}
Chemin, L., Carignan, C., \& Foster, T. 2009, ApJ, 705, 1395

\bibitem[{Chy\.zy(2008)}]{Chyzy:2008a}
Chy\.zy, K.~T. 2008, A\&A, 482, 755

\bibitem[{Chy\.zy et~al.(2000)Chy\.zy, Beck, Kohle, Klein, \&
  Urbanik}]{Chyzy:2000}
Chy\.zy, K.~T., Beck, R., Kohle, S., Klein, U., \& Urbanik, M. 2000, A\&A, 355,
  128

\bibitem[{Chy\.zy \& Buta(2008)}]{Chyzy:2008b}
Chy\.zy, K.~T., \& Buta, R.~J. 2008, ApJ, 677, L17

\bibitem[{Ehle \& Beck(1993)}]{Ehle:1993}
Ehle, M., \& Beck, R. 1993, A\&A, 273, 45

\bibitem[{Fitt \& Alexander(1993)}]{Fitt:1993}
Fitt, A.~J., \& Alexander, P. 1993, MNRAS, 261, 445

\bibitem[{Fletcher et~al.(2010)Fletcher, Beck, Shukurov, Berkhuijsen, \&
  Horellou}]{Fletcher:2010}
Fletcher, A., Beck, R., Shukurov, A., Berkhuijsen, E.~M., \& Horellou, C. 2010,
  MNRAS, submitted. (arXiv:1001.5230)

\bibitem[{Fletcher et~al.(2004)Fletcher, Berkhuijsen, Beck, \&
  Shukurov}]{Fletcher:2004}
Fletcher, A., Berkhuijsen, E.~M., Beck, R., \& Shukurov, A. 2004, A\&A, 414, 53

\bibitem[{Frick et~al.(2000)Frick, Beck, Shukurov, Sokoloff, Ehle, \&
  Kamphuis}]{Frick:2000}
Frick, P., Beck, R., Shukurov, A., Sokoloff, D., Ehle, M., \& Kamphuis, J.
  2000, MNRAS, 318, 925

\bibitem[{Gaensler et~al.(2004)Gaensler, Beck, \& Feretti}]{Gaensler:2004}
Gaensler, B.~M., Beck, R., \& Feretti, L. 2004, New Astronomy Reviews, 48, 1289

\bibitem[{Gaensler et~al.(2005)Gaensler, Haverkorn, Staveley-Smith, Dickey,
  McClure-Griffiths, Dickel, \& Wolleben}]{Gaensler:2005}
Gaensler, B.~M., Haverkorn, M., Staveley-Smith, L., Dickey, J.~M.,
  McClure-Griffiths, N.~M., Dickel, J.~R., \& Wolleben, M. 2005, Science, 307,
  1610

\bibitem[{Garcia-Burillo et~al.(1993)Garcia-Burillo, Combes, \&
  Gerin}]{Garcia-Burillo:1993}
Garcia-Burillo, S., Combes, F., \& Gerin, M. 1993, A\&A, 274, 148

\bibitem[{Gissinger et~al.(2009)Gissinger, Fromang, \& Dormy}]{Gissinger:2009}
Gissinger, C., Fromang, S., \& Dormy, E. 2009, MNRAS, 394, L84

\bibitem[{Gressel et~al.(2008)Gressel, Elstner, Ziegler, \&
  R{\"u}diger}]{Gressel:2008}
Gressel, O., Elstner, D., Ziegler, U., \& R{\"u}diger, G. 2008, A\&A, 486, L35

\bibitem[{Han et~al.(1998)Han, Beck, \& Berkhuijsen}]{Han:1998}
Han, J.~L., Beck, R., \& Berkhuijsen, E.~M. 1998, A\&A, 335, 1117

\bibitem[{Han et~al.(1999)Han, Beck, Ehle, Haynes, \& Wielebinski}]{Han:1999}
Han, J.~L., Beck, R., Ehle, M., Haynes, R.~F., \& Wielebinski, R. 1999, A\&A,
  348, 405

\bibitem[{Heesen et~al.(2009)Heesen, Krause, Beck, \& Dettmar}]{Heesen:2009}
Heesen, V., Krause, M., Beck, R., \& Dettmar, R.-J. 2009, A\&A, 506, 1123

\bibitem[{Hummel(1986)}]{Hummel:1986}
Hummel, E. 1986, A\&A, 160, L4

\bibitem[{Kepley et~al.(2010)Kepley, Muhle, Everett, Zweibel, Wilcots, \&
  Klein}]{Kepley:2010}
Kepley, A.~A., Muhle, S., Everett, J., Zweibel, E.~G., Wilcots, E.~M., \&
  Klein, U. 2010, ApJ, 712, 536

\bibitem[{Krasheninnikova et~al.(1989)Krasheninnikova, Shukurov, Ruzmaikin, \&
  Sokolov}]{Krasheninnikova:1989}
Krasheninnikova, I., Shukurov, A., Ruzmaikin, A., \& Sokolov, D. 1989, A\&A,
  213, 19

\bibitem[{Krause \& Raedler(1980)}]{Krause:1980}
Krause, F., \& Raedler, K.-H. 1980, Mean-field magnetohydrodynamics and dynamo
  theory (Oxford: Pergamon Press)

\bibitem[{Krause(2009)}]{Krause:2009}
Krause, M. 2009, in Magnetic Fields in the Universe II: From Laboratory and
  Stars to the Primordial Universe, edited by A.~Esquivel, J.~Franco,
  G.~Garcia-Segura, E.~M. de~Gouveia Dal~Pino, A.~Lazarian, S.~Lizano, \&
  A.~Raga (UNAM, Mexico), vol.~36 of Rev. Mex. AA, 25

\bibitem[{Krause et~al.(1989{\natexlab{a}})Krause, Beck, \&
  Hummel}]{Krause:1989b}
Krause, M., Beck, R., \& Hummel, R. 1989{\natexlab{a}}, A\&A, 217, 17

\bibitem[{Krause et~al.(1989{\natexlab{b}})Krause, Hummel, \&
  Beck}]{Krause:1989a}
Krause, M., Hummel, E., \& Beck, R. 1989{\natexlab{b}}, A\&A, 217, 4

\bibitem[{Kulsrud \& Zweibel(2008)}]{Kulsrud:2008}
Kulsrud, R.~M., \& Zweibel, E.~G. 2008, Rep. Prog. Phys., 71, 1

\bibitem[{Mao et~al.(2008)Mao, Gaensler, Stanimirovic, Haverkorn,
  McClure-Griffiths, Staveley-Smith, \& Dickey}]{Mao:2008}
Mao, S.~A., Gaensler, B.~M., Stanimirovic, S., Haverkorn, M.,
  McClure-Griffiths, N.~M., Staveley-Smith, L., \& Dickey, J.~M. 2008, ApJ,
  688, 1029

\bibitem[{Moss et~al.(1998)Moss, Shukurov, Sokoloff, Berkhuijsen, \&
  Beck}]{Moss:1998}
Moss, D., Shukurov, A., Sokoloff, D.~D., Berkhuijsen, E.~M., \& Beck, R. 1998,
  A\&A, 335, 500

\bibitem[{Moss et~al.(2007)Moss, Snodin, Englmaier, Shukurov, Beck, \&
  Sokoloff}]{Moss:2007}
Moss, D., Snodin, A.~P., Englmaier, P., Shukurov, A., Beck, R., \& Sokoloff,
  D.~D. 2007, A\&A, 465, 157

\bibitem[{Neininger(1992)}]{Neininger:1992}
Neininger, N. 1992, A\&A, 263, 30

\bibitem[{Niklas(1995)}]{Niklas:1995}
Niklas, S. 1995, Ph.D. thesis, University of Bonn

\bibitem[{Niklas et~al.(1997)Niklas, Klein, \& Wielebinski}]{Niklas:1997}
Niklas, S., Klein, U., \& Wielebinski, R. 1997, A\&A, 322, 19

\bibitem[{Patrikeev et~al.(2006)Patrikeev, Fletcher, Stepanov, Beck,
  Berkhuijsen, Frick, \& Horellou}]{Patrikeev:2006}
Patrikeev, I., Fletcher, A., Stepanov, R., Beck, R., Berkhuijsen, E.~M., Frick,
  P., \& Horellou, C. 2006, A\&A, 458, 441

\bibitem[{Poezd et~al.(1993)Poezd, Shukurov, \& Sokoloff}]{Poezd:1993}
Poezd, A., Shukurov, A., \& Sokoloff, D. 1993, MNRAS, 264, 285

\bibitem[{Rohde et~al.(1999)Rohde, Beck, \& Elstner}]{Rohde:1999}
Rohde, R., Beck, R., \& Elstner, D. 1999, A\&A, 350, 423

\bibitem[{Ruzmaikin et~al.(1988)Ruzmaikin, Shukurov, \&
  Sokoloff}]{Ruzmaikin:1988}
Ruzmaikin, A.~A., Shukurov, A.~M., \& Sokoloff, D.~D. 1988, Magnetic Fields of
  Galaxies (Dordrecht: Kluwer)

\bibitem[{Shukurov(2007)}]{Shukurov:2007}
Shukurov, A. 2007, in Mathematical Aspects of Natural Dynamos, edited by D.~E.,
  \& S.~A. (London: Chapman \& Hall/CRC), vol.~13 of The Fluid Mechanics of
  Astrophysics and Geophysics, 319

\bibitem[{Shukurov et~al.(2006)Shukurov, Sokoloff, Subramanian, \&
  Brandenburg}]{Shukurov:2006}
Shukurov, A., Sokoloff, D., Subramanian, K., \& Brandenburg, A. 2006, A\&A,
  448, L33

\bibitem[{Soida et~al.(2002)Soida, Beck, Urbanik, \& Braine}]{Soida:2002}
Soida, M., Beck, R., Urbanik, M., \& Braine, J. 2002, A\&A, 394, 47

\bibitem[{Sokoloff et~al.(1998)Sokoloff, Bykov, Shukurov, Berkhuijsen, Beck, \&
  Poezd}]{Sokoloff:1998}
Sokoloff, D.~D., Bykov, A.~A., Shukurov, A., Berkhuijsen, E.~M., Beck, R., \&
  Poezd, A.~D. 1998, MNRAS, 299, 189

\bibitem[{Stepanov et~al.(2008)Stepanov, Arshakian, Beck, Frick, \&
  Krause}]{Stepanov:2008}
Stepanov, R., Arshakian, T.~G., Beck, R., Frick, P., \& Krause, M. 2008, A\&A,
  480, 45

\bibitem[{Strong et~al.(2000)Strong, Moskalenko, \& Reimer}]{Strong:2000}
Strong, A.~W., Moskalenko, I.~V., \& Reimer, O. 2000, ApJ, 537, 763

\bibitem[{Sur et~al.(2007)Sur, Shukurov, \& Subramanian}]{Sur:2007}
Sur, S., Shukurov, A., \& Subramanian, K. 2007, MNRAS, 377, 874

\bibitem[{Tabatabaei et~al.(2007)Tabatabaei, Beck, Krause, Berkhuijsen, Gordon,
  \& Menten}]{Tabatabaei:2007}
Tabatabaei, F., Beck, R., Krause, M., Berkhuijsen, E.~M., Gordon, K.~D., \&
  Menten, K.~M. 2007, A\&A, 475, 133

\bibitem[{Tabatabaei et~al.(2008)Tabatabaei, Krause, Fletcher, \&
  Beck}]{Tabatabaei:2008}
Tabatabaei, F., Krause, M., Fletcher, A., \& Beck, R. 2008, A\&A, 490, 1005

\bibitem[{Widrow(2002)}]{Widrow:2002}
Widrow, L.~M. 2002, Rev. Mod. Phys., 74, 775

\end{thebibliography}

\end{document}